\documentclass[10pt,twocolumn,a4paper]{esaAI}
\usepackage[dvipsnames]{xcolor}
\usepackage{comment}
\title{Heterogeneity: An Open Challenge for Federated On-board Machine Learning}

\def\authorEmail{\{firstname.lastname\}@uni.lu}

\author[1]{Maria Hartmann\thanks{Corresponding author. E-Mail: \authorEmail}}
\author[1,2]{Gr{\'e}goire Danoy}
\author[2]{Pascal Bouvry}
\affil[1]{SnT, University of Luxembourg, Luxembourg}
\affil[2]{FSTM, University of Luxembourg, Luxembourg}

\begin{document}

\makeCustomtitle

\begin{abstract}
The design of satellite missions is currently undergoing a paradigm shift from the historical approach of individualised monolithic satellites towards distributed mission configurations, consisting of multiple small satellites. With a rapidly growing number of such satellites now deployed in orbit, each collecting large amounts of data, interest in on-board orbital edge computing is rising. Federated Learning is a promising distributed computing approach in this context, allowing multiple satellites to collaborate efficiently in training on-board machine learning models. Though recent works on the use of Federated Learning in orbital edge computing have focused largely on homogeneous satellite constellations, Federated Learning could also be employed to allow heterogeneous satellites to form ad-hoc collaborations, e.g.~ in the case of communications satellites operated by different providers. Such an application presents additional challenges to the Federated Learning paradigm, arising largely from the heterogeneity of such a system. In this position paper, we offer a systematic review of these challenges in the context of the cross-provider use case, giving a brief overview of the state-of-the-art for each, and providing an entry point for deeper exploration of each issue. 
\end{abstract}

\section{Introduction}
\begin{figure}[th]
    \centering
    \vspace{-2em}
    \includegraphics[width=.99\columnwidth]{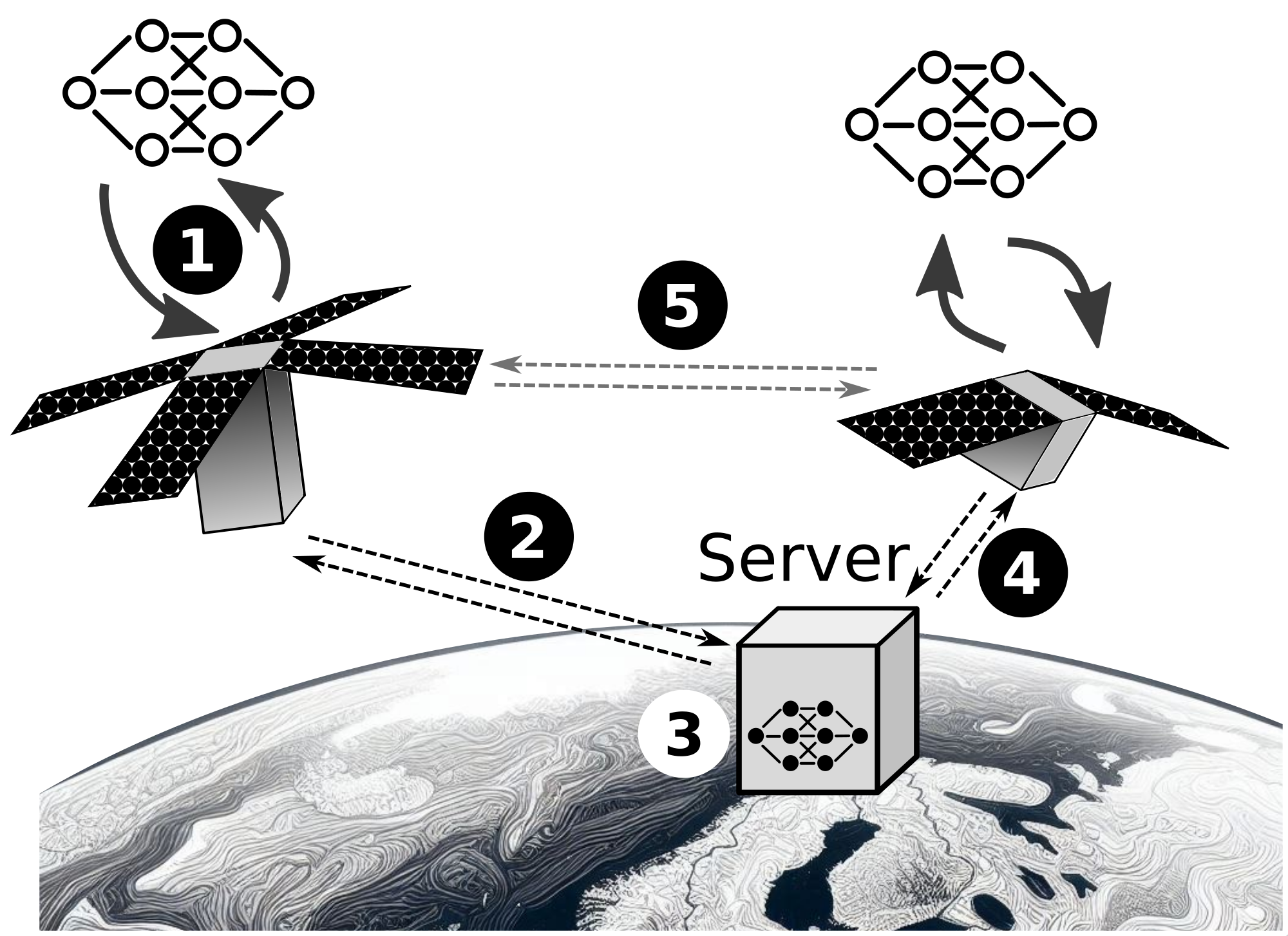}
	\caption{In Federated Learning, each satellite performs on-board machine learning to train a local model (1). Only these models are transmitted via satellite link to a server (2), here based on the ground, where multiple local models are aggregated into a single global model (3). This global model is transmitted back to the satellites (4) to continue the learning process. If necessary, satellites can act as relays for one another (5).
	}
	\label{fig:fl-heterogeneous}
\end{figure}

With advances in hardware and software capabilities, distributed satellite mission configurations are progressively replacing the classical paradigm of using a single monolithic spacecraft. 
With nanosatellites able to generate and store increasingly large amounts of data through various on-board sensors, downlink capacity is becoming a major bottleneck in processing the gathered information. 
To manage this problem, there is an ongoing drive towards shifting data processing onto satellites\cite{izzo2022selected} -- this strategy is referred to as Orbital Edge Computing (OEC)\cite{Denby2020}. The overarching idea of OEC is to leverage on-board computing capabilities of each satellite to process locally gathered data, reducing the size and amount of required transmissions and speeding up evaluation. 
A promising variant of OEC proposes deploying Federated Learning~\cite{McMahan2017} (FL) on satellites, allowing the joint training of on-board machine learning models across the data gathered by multiple satellites with a limited communication budget~\cite{Jabbarpour2023}. Under a FL scheme, each satellite performs on-board machine learning on the data it collects, training a local model -- see Figure~\ref{fig:fl-heterogeneous}. These models are shared periodically among participants, allowing them to be aggregated into a more accurate global model on which to continue training. Aggregation can take place with the aid of a parameter server on the ground or in orbit, or in a fully decentralised manner between satellites. Fundamental advantages of this approach include a vastly reduced communication cost compared to the transmission of raw data, and the inherent privacy advantages of compartmentalising data on satellites.\\
Current literature on the use of Federated Learning in Orbital Edge Computing is focused primarily on a single use case: using Federated Learning in a single, dedicated constellation of satellites. However, another frequently occurring scenario appears largely unstudied: the potential for satellites from different missions and providers to form (ad-hoc) cross-provider collaborations.
In this position paper, we offer an initial exploration of the conceptual challenges associated with this use case: we identify the characteristics of the problem, present a brief survey of the state of the art for each, connecting existing research from the application domain and the theoretical field, and discuss how existing approaches might fare in this scenario. We conclude with a gap analysis.

\section{Survey results}

In this section, we analyse the current state of the art in research relating to our use case of cross-provider FL. As this use case has not yet been addressed explicitly, we divide our analysis into different thematic sections. We begin by considering the research closest to application, considering orbital edge computing (OEC) and Federated learning (FL) schemes tailored to use on satellites. Following this, we discuss cross-provider FL, an OEC use case that has, to the best of our knowledge, not been addressed to date. We explore related research from the field of FL that could be applicable for this use case, particularly works addressing the handling of different types of heterogeneity, and discuss their applicability.%

\subsection{Orbital edge computing and federated learning}
Federated Learning could offer a flexible framework for satellites to collaborate on on-board information processing\cite{izzo2022selected}\cite{Chen2022} while limiting communication cost and preserving data privacy.
Various works modify the FL paradigm for the use case of LEO constellations, mainly focusing on adapted communication schedules to handle the intermittent connectivity of satellites. These approaches can broadly be divided by their proposed %
placement of a parameter server.

Initial works focused on the use of a ground-based server, offering a relatively higher resource capacity of the server; the drawback is a communication bottleneck caused by the intermittent connectivity of satellites. 
The FedSpace algorithm~\cite{So2022} attempts to overcome this constraint by performing semi-asynchronous federated aggregation, exploiting knowledge about clients' orbital periods to calculate an aggregation schedule that yields an optimal trade-off between satellite idleness and model staleness.%
The FedGSM algorithm\cite{Wu2023} similarly makes use of known connectivity intervals to extrapolate model updates.

Conversely, FedISL\cite{Razmi2022}, a synchronous FL scheme for a dense LEO constellation, hinges on the strategic placement of a server in medium Earth orbit (MEO); with convergence speed further enhanced by the use of intra-plane inter-satellite links.
This concept is extended in~\cite{Razmi2024} with the grouping of satellites sharing the same orbit to speed up aggregation. Similarly, the synchronous FedHAP~\cite{Elmahallawy2022} algorithm is based on the deployment of multiple high-altitude aerial platforms to accelerate aggregation; AsyncFLEO~\cite{Elmahallawy2022b} rests on the same premise, but is capable of asynchronous aggregation.
The DSFL\cite{Wu2022} algorithm side-steps the challenge of server placement by performing fully decentralised aggregation. Finally, in their work on semi-supervised FL, \"{O}stman et al.~\cite{Ostman2023} compare a decentralised aggregation strategy with two scenarios where a relay satellite and a set of ground stations, respectively, act as the aggregation server. All three variants are shown to achieve comparable accuracy performances with similar total training time and power consumption.%

Note that the works presented in this section do not consider heterogeneity challenges in great depth; several works, e.g.~\cite{So2022}\cite{Razmi2024}, claim that any suitable FL algorithm could be utilised as a drop-in component. In the following section, we assess the additional characteristics that might be required of an algorithm to mitigate heterogeneity in the cross-provider use case. 
\subsection{Heterogeneity challenges of cross-provider FL}
With the proliferation of private and commercial missions, an ever-increasing number of satellites with different capabilities and overlapping interests are active in Earth orbit. %
Enabling an ad-hoc collaboration between satellites of providers with compatible interests could serve to enhance the performance of all sides %
at a comparatively low communication cost.
A natural example is the application to satellite communication problems: %
machine learning has the potential to assist with various SatComm-related problems\cite{Fourati2021}, and collaboration between satellites could help solve these problems with greater accuracy and reliability. %
Many such satellites by different service providers are in operation today, with different hardware, different orbits and different underlying purposes, but nevertheless carrying out related functions. %
Compared to performing FL on single-mission satellite constellations, %
this application scenario presents unique challenges induced by the heterogeneity of satellites.

This could involve many different types of heterogeneity, known to present a challenge to FL algorithms\cite{Kairouz2021}. These include \textit{data heterogeneity}, \textit{feature heterogeneity}, %
\textit{device heterogeneity}, %
and \textit{preference heterogeneity}. %
In this section, we discuss the state-of-the-art approaches for each of these types, highlighting how each has been addressed for the OEC use case and, where missing, how existing solutions might transfer to this use case. 

\textbf{Data heterogeneity.} This type of heterogeneity, where data is imbalanced across participants, is discussed extensively in the literature\cite{Kairouz2021}, as it occurs naturally in most settings. In our use case, heterogeneous distributions of data across satellites are quite likely, with the extent dependent in part on the precise setting. For example, satellites gathering Earth observation images might collect significantly different samples based on their orbital planes, while for SatComm data might differ based on the role of the satellite or the associated service provider. The general issue of data heterogeneity is discussed in most FL variants proposed for the OEC use case, e.g.~\cite{Wu2022}, \cite{Wu2023}; however, their effectiveness is seldom demonstrated beyond preliminary benchmarking experiments.
Therefore, it appears worthwhile to also consider the state of the art in the general field of FL. A taxonomy of variants of data heterogeneity is presented in \cite{Ye2023}, along with a comprehensive survey of current mitigation approaches. According to \cite{Ye2023}, these can be broadly divided into data-level, model-level and server-level interventions. Data-level approaches involve modifying the underlying training data to balance heterogeneity, e.g.~by preprocessing~\cite{Li2021}\cite{Xu2023}, generating supplemental data using Generative Adversarial Networks~\cite{Goodfellow2020}, or transmitting information about data between clients~\cite{Yoon2021FedMixAO}. However, these strategies often place a significant additional computing or communication burden on the clients, rendering them unattractive use on satellites. Selected model- and server-level strategies appear more promising, as they either require little additional computation cost, or can be carried out on the server-side. Notably, these include model regularisation~\cite{Kim2022MultiLevelBR}, knowledge distillation~\cite{Zhu2021}, and personalised federated learning (PFL) approaches~\cite{Tan2023} such as client clustering\cite{Ghosh2022}\cite{Duan2021}, parameter decoupling~\cite{Arivazhagan2019} and model interpolation~\cite{Hanzely2020FederatedLO}.

It is difficult to single out an optimal approach for the general version of our use case, where the data distribution pattern is unknown. %
The most promising approach would likely be an adaptive solution that modifies the aggregation approach during runtime based on observed metrics, e.g. clustering participants by similarity~\cite{Ghosh2022}\cite{Duan2021} or assigning importance weights for aggregation~\cite{Hanzely2020FederatedLO}. For mission architectures involving a powerful ground-based parameter server, more complex knowledge distillation-based approaches may also be an option.

Finally, we note that if the nature of the data distribution is known, such as in EO imaging missions, this could be exploited to the advantage of the algorithm, e.g.~by grouping participants known to collect similar data, or conversely by exchanging small sets of selected samples to balance highly different datasets.

\textbf{Device heterogeneity.}
Aside from the communication challenges induced by orbital trajectories, satellites in the cross-provider setting would also%
have hardware differences, leading to different levels of noise and training datasets of varying quality, and impacting computational speeds and capabilities. This is a common problem in the general field of FL~\cite{Ye2023}; standard approaches include adaptively assigning different weight contributions~\cite{Ma2022} or model architectures~\cite{Diao2020HeteroFLCA} to participants; possibly also reducing the consideration of lower-quality participants in selecting clients for aggregation~\cite{Li2021b}\cite{Nishio2019}. It appears likely that such strategies would transfer well to the present use case, without a need for major modifications. %

\textbf{Feature heterogeneity.} This setting corresponds to the collaboration of satellites with different types of sensors, collecting different types of data and potentially requiring different model architectures for on-board processing. Effectively integrating different features and models into a coherent federated model training process presents a difficult problem; to the best of our knowledge, it has not yet been tackled in research on OEC. %
Indeed, the general problem of performing FL in such a setting, known as Vertical Federated Learning (VFL)~\cite{Yang2019}, remains largely unsolved beyond highly constrained artificial scenarios or costly compensation approaches~\cite{Liu2024}. This present lack of solutions renders the application of FL to the real-world satellite use case infeasible. The only approach that appears viable at present would involve feature distillation on a ground-station parameter server -- a computationally expensive solution that does not yield workable models for the federated participants, and so would benefit only the server-side~\cite{Liu2024}. Beyond this niche variant, the currently most feasible approach would likely be to separate participants by features, eliminating this type of heterogeneity. 

\textbf{Preference heterogeneity.} This is a novel type of heterogeneity that, as of now, has seen little recognition in the field of federated learning; yet it appears highly relevant to the present use case. Preference heterogeneity arises when participants are solving problems with multiple objectives, e.g.~minimising communication cost while also minimising connection latency in a satellite communication network. In such multi-objective problems, different optimal solutions are generally possible, representing different trade-offs between the individual objectives. In practice, some trade-offs may be more desirable than others, e.g.~conserving energy by limiting communication for severely resource-constrained satellites, or minimising connection latency to boost service to certain geographical areas. This can be controlled by assigning importance weights to each objective when solving the learning problem. We call participants with different importance weights preference-heterogeneous.

At present, little research has been devoted to performing Federated Learning under preference heterogeneity; the closest related works consider federated multi-task learning~\cite{Smith2017}, where participants have fully separate objectives, and federated multi-objective learning~\cite{yang_federated_2023}\cite{hartmann2023mofld} without allowance for different preferences. 

\subsection{Other considerations}
\subsubsection{Fairness} 
In addition to the technical considerations of the previous section, this cross-provider use case also differs from the 
single-provider variant in the assumptions made about participants' intentions. Satellites designed to collaborate with each other as part of a single mission can generally be assumed to act altruistically in federation, i.e.~towards the benefit of the larger system. In the cross-provider setting, however, no such assumption should be made, as has been noted e.g.~in \cite{Razmi2024}.
Instead, we assume that participants may act `selfishly', valuing their own success over that of the federated system. This could occur e.g.~ if satellites contribute low-quality updates, leading to an overall degradation of the global model and benefiting from the collaboration at the cost of others. Similarly, a participant could limit the frequency of its contributions to conserve communication budget or maintain privacy, to the detriment of others in the federation, while still receiving global model updates. A successful cross-provider collaboration scheme should guard against such exploitation. These fairness considerations have yet to be addressed in works targeting the OEC use case; a full survey on the state of the art of general FL approaches is provided in~\cite{Shi2024}. The same work gives a detailed account of the different definitions of fairness and underlying assumptions; the choice of an appropriate mitigation approach for the present depends on these characteristics.
\subsubsection{Guarding against malicious participants}
This is a more extreme case of the challenges discussed in the previous section -- here participants intentionally attempt to sabotage the performance of the federated system through their participation, e.g.~through submitting intentionally false model updates (known as model poisoning attacks). A thorough overview of possible attack vectors and approaches for guarding against such attacks is given in \cite{Kairouz2021} and more recently \cite{RodrguezBarroso2023}, suggesting e.g.~the assignment of confidence scores~\cite{MunozGonzlez2019ByzantineRobustFM}, filtering outliers~\cite{Yin2018}, or normalising model updates before aggregation~\cite{Sun2019}. A number of these solutions appear to transfer well to the cross-provider OEC use case, given a trustworthy server, with the selection of approach dependent on the particular parameters of the system and the results of a risk analysis. 

\subsubsection{The role of standardisation}
Along with the algorithmic approaches towards preventing misuse of the collaboration, standardisation likely has an important role to play in the deployment of FL to the present use case involving multiple providers. This could e.g.~ include a requirement for certification of machine learning pipelines in accordance with certain quality standards to obtain access to such federated exchange schemes, to decrease the likelihood of interference by malicious or poorly engineered participants. A review of standards relating to the trustworthiness of machine learning for space applications suggests that such standards largely have yet to be defined~\cite{ILNASWP1}; a recently published handbook provides a first glimpse of such considerations~\cite{MLHandbookECSS}.

Finally, we note another crucial challenge of this particular use case: the need for a unified communication protocol between participants, capable of negotiating the parameters of the FL scheme between participants and able to transmit machine learning models unambiguously, without interpretation errors caused e.g.~by differences in hardware or software. For ad-hoc collaboration, standardised communication formats are of critical importance, both to negotiate the parameters of the FL protocol during the initialisation phase, and to transmit model updates without error. There appears to be no existing solution, as most works in the literature tend to assume a group of satellites collaborating as part of a single unified mission.

\section{Discussion}
In our analysis of the state of the art, we have seen that current works investigating the use of Federated Learning in satellite on-board edge computing have largely focused on a single scenario: a large constellation of homogeneous satellites deployed as part of a single mission. We have introduced a different use case, relevant to the present or near future, where heterogeneous satellites of multiple different providers could collaborate under a FL scheme to enhance on-board learning. We have elucidated the unique conceptual challenges of this use case, with a focus on the different types of heterogeneity and conflicts of interest that might arise. We have provided a broad perspective of the state of the art for each, discussing both the existing work close to the use case, and the general state of the art in the theoretical literature. Our brief survey shows that several aspects remain to be addressed to adequately solve this real-world use case. In particular, there is a need to further investigate (1) how state-of-the-art solutions can be combined in settings where multiple types of heterogeneity occur simultaneously; (2) which heterogeneity-mitigating algorithms could be selected to fit with the use case, and with existing OEC schemes; (3) how to perform Federated Learning under preference heterogeneity. Finally, we suggest that %
any engineering approach should be supported by additional measures reducing the complexity of the system by providing appropriate constraints, e.g.~through the use of standardisation.

\section*{Acknowledgements} %
This work is partially funded by the joint research programme UL/SnT--ILNAS on Technical Standardisation for Trustworthy ICT, Aerospace, and Construction.

\printbibliography
\addcontentsline{toc}{section}{References}

\end{document}